\documentclass[sigconf]{acmart}

\AtBeginDocument{%
  }

\setcopyright{acmlicensed}
\copyrightyear{2018}
\acmYear{2018}
\acmDOI{XXXXXXX.XXXXXXX}

\acmConference[CIKM '24]{Make sure to enter the correct
  conference title from your rights confirmation emai}{October 21--25,
  2024}{Boise, Idaho, USA}
\acmISBN{978-1-4503-XXXX-X/18/06}

\usepackage{algpseudocode}
\usepackage{algorithm}
\usepackage{xcolor}
\usepackage{enumitem} 
\usepackage{amsfonts} 
\usepackage{tabularx}
\usepackage{array}
\usepackage{booktabs}
\usepackage{amsmath}
\usepackage{multirow} 
\usepackage{siunitx}
\usepackage{graphicx}
\usepackage{subcaption}
\usepackage{caption}
\usepackage{booktabs} 
\usepackage{siunitx} 
\copyrightyear{2024}
\acmYear{2024}
\setcopyright{acmlicensed}\acmConference[CIKM '24]{Proceedings of the 33rd ACM International Conference on Information and Knowledge Management}{October 21--25, 2024}{Boise, ID, USA}
\acmBooktitle{Proceedings of the 33rd ACM International Conference on Information and Knowledge Management (CIKM '24), October 21--25, 2024, Boise, ID, USA}
\acmDOI{10.1145/3627673.3679756}
\acmISBN{979-8-4007-0436-9/24/10}

\newtheorem{lemma}{Lemma}

\begin{document}

\newcolumntype{L}[1]{>{\raggedright\arraybackslash}p{#1}}
\newcolumntype{C}[1]{>{\centering\arraybackslash}p{#1}}
\newcolumntype{R}[1]{>{\raggedleft\arraybackslash}p{#1}}
\title{Bridging User Dynamics: Transforming Sequential Recommendations with Schrödinger Bridge and Diffusion Models}

\author{Wenjia Xie}
\email{xiaohulu@mail.ustc.edu.cn}
\orcid{0009-0009-6392-2559}
\affiliation{%
  \institution{University of Science and Technology of China \& State Key Laboratory of Cognitive Intelligence}
  \city{Hefei}
  \country{China}
}

\author{Rui Zhou}
\email{zhou_rui@mail.ustc.edu.cn}
\orcid{0009-0005-9561-7496}
\affiliation{%
  \institution{University of Science and Technology of China \& State Key Laboratory of Cognitive Intelligence}
  \city{Hefei}
  \country{China}
}

\author{Hao Wang}
\authornote{Corresponding author.}
\email{wanghao3@ustc.edu.cn}
\orcid{0000-0001-9921-2078}
\affiliation{%
  \institution{University of Science and Technology of China \& State Key Laboratory of Cognitive Intelligence}
  \city{Hefei}
  \country{China}
}

\author{Tingjia Shen}
\email{jts_stj@mail.ustc.edu.cn}
\orcid{0009-0003-5949-7218}
\affiliation{%
  \institution{University of Science and Technology of China \& State Key Laboratory of Cognitive Intelligence}
  \city{Hefei}
  \country{China}
}

\author{Enhong Chen}
\email{cheneh@ustc.edu.cn}
\orcid{0000-0002-4835-4102}
\affiliation{%
  \institution{University of Science and Technology of China \& State Key Laboratory of Cognitive Intelligence}
  \city{Hefei}
  \country{China}
}

\renewcommand{\shortauthors}{Wenjia Xie, Rui Zhou, Hao Wang, Tingjia Shen, Enhong Chen}

\begin{abstract}
 
Sequential recommendation has attracted increasing attention due to its ability to accurately capture the dynamic changes in user interests. We have noticed that generative models, especially diffusion models, which have achieved significant results in fields like image and audio, hold considerable promise in the field of sequential recommendation. However, existing sequential recommendation methods based on diffusion models are constrained by a prior distribution limited to Gaussian distribution, hindering the possibility of introducing user-specific information for each recommendation and leading to information loss.
To address these issues, we introduce the Schrödinger Bridge into diffusion-based sequential recommendation models, creating the SdifRec model. This allows us to replace the Gaussian prior of the diffusion model with the user's current state, directly modeling the process from a user's current state to the target recommendation. Additionally, to better utilize collaborative information in recommendations, we propose an extended version of SdifRec called con-SdifRec, which utilizes user clustering information as a guiding condition to further enhance the posterior distribution. Finally, extensive experiments on multiple public benchmark datasets have demonstrated the effectiveness of SdifRec and con-SdifRec through comparison with several state-of-the-art methods. Further in-depth analysis has validated their efficiency and robustness. 
\vspace{-2mm}
\end{abstract}

\ccsdesc[500]{Information systems~Recommender systems}
\keywords{Sequential Recommendation, Schrödinger Bridge, Diffusion Model, Classifier-free Guidance}
\maketitle

\section{Introduction}

In recent years, due to the outstanding performance and significant business value, sequential recommendation (SR) has attracted increasing attention~\cite{chen2020sequence,rendle2010factorizing,yin2024dataset, han2024end4rec}.
Distinct from the traditional collaborative filtering or certain graph-based approaches, SR systems underscore the dynamic behaviors inherent to users themselves, rather than depending solely on structured data~\cite{chen2022intent,wang2019mcne}. This confers enhanced personalization and its ability to more precisely track the shifts in users' interests and demands. Prominent deep learning-based SR models utilize the CNN, RNN, and GNN architecture to model users' preferences from historical interaction records, such as
Caser~\cite{tang2018personalized}, GRU4Rec~\cite{hidasi2015session}, and SR-GNN~\cite{DBLP:journals/corr/abs-1811-00855}. After that, SASRec~\cite{kang2018self} has been a pioneering work that introduces Transformer~\cite{c:22} into SR to capture dependencies with powerful modeling capability. BERT4Rec~\cite{sun2019bert4rec} further adopts BERT architecture~\cite{devlin2018bert} and utilizes a masked language model to predict the target item.

\begin{figure}[htbp] 
\centering 
\begin{subfigure}{\columnwidth} 
\includegraphics[width=\linewidth]{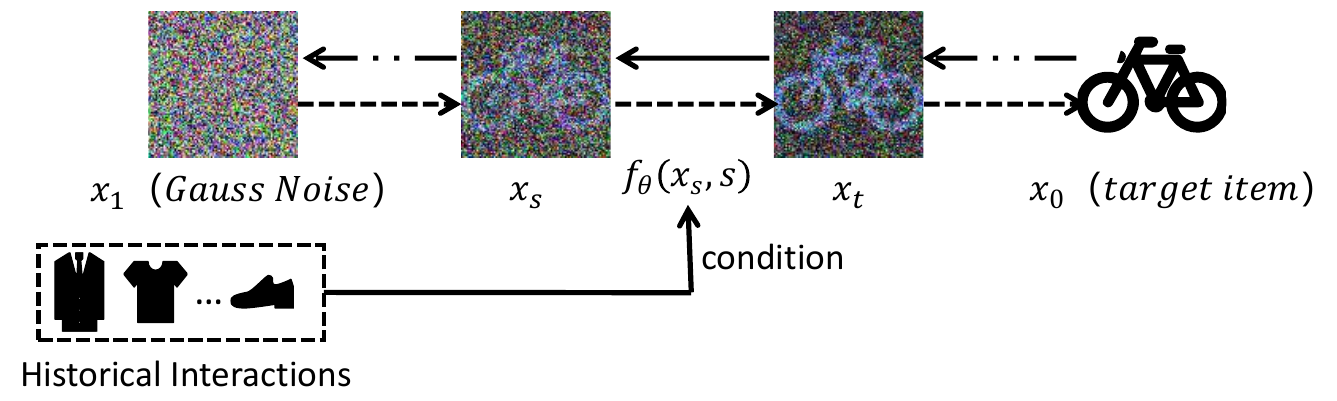} 
\caption{Typically diffusion-based sequential recommendation method.}
\label{fig:sub1}
\end{subfigure}

\begin{subfigure}{\columnwidth} 
\includegraphics[width=\linewidth]{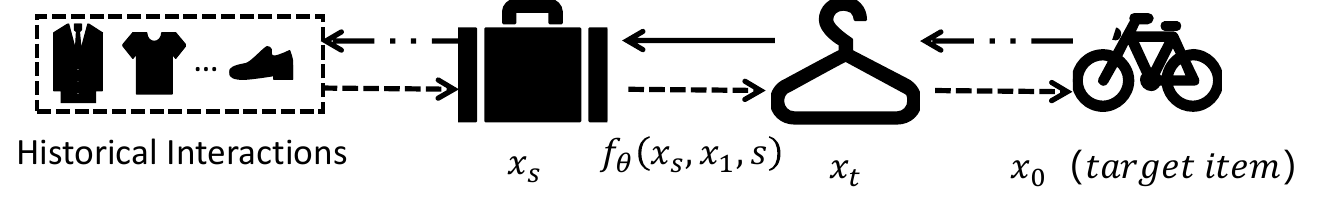} 
\vspace{-6mm}
\caption{Replace the Gaussian noise with historical interaction information as the prior distribution.}
\label{fig:sub2}
\end{subfigure}
\vspace{-6mm}
\caption{An example illustrates the difference between our motivation and existing diffusion-based methods.}
\label{fig:motivation}

\vspace{-6mm}
\end{figure}

With the rapid development of generative models, some studies have applied them to SR and achieved significant improvements. For example, SVAE~\cite{sachdeva2019sequential} effectively models the probability distribution of the most likely future preferences by combining variational autoencoders (VAE)~\cite{kingma2013auto} and GRU~\cite{cho2014learning}. MFGAN~\cite{ren2020sequential} decouples factors in SR based on the Generative Adversarial Network (GAN)~\cite{goodfellow2020generative} and trains the model using policy gradients. 
However, these methods are constrained by the expressive power and generative quality of VAE and GAN themselves~\cite{alemi2016deep,salimans2016improved} and face the issue of posterior collapse~\cite{lucas2019understanding}, where the generated hidden representations often lack critical information about user preferences. 
As a result, we have turned our attention to a new paradigm of generative models, diffusion models~\cite{ho2020denoising}, which have recently achieved exciting achievements in fields such as image and text generation~\cite{dhariwal2021diffusion, li2022diffusionlm}. There have already been a few works based on diffusion models in SR, and they have achieved satisfactory results, such as DiffRec~\cite{wang2023diffusion} and Diffurec~\cite{li2023diffurec}. These methods follow the principles of diffusion models, initially perturbing the embedding of the target item through a forward diffusion process into a known prior distribution, that is Gaussian noise.  Subsequently, they restore the Gaussian distribution iteratively through a reverse denoising process, also referred to as the sampling phase, to recover meaningful representations and recommend items that are most similar to it.

However, adhering to the paradigm of diffusion models, the prior distribution of these SR methods based on diffusion models is confined to a Gaussian distribution. Thus they can only utilize historical interactions as conditional information for the model. This constrains the potential of diffusion models, as only target items undergo the diffusion model processing. 
Additionally, information in SR is often sparse yet crucial~\cite{he2016fusing}; during the process of adding noise to the pure noise state, the information is further compromised, making the model prone to collapse.
 Therefore, we aim to modify the diffusion model by substituting the Gaussian prior with meaningful historical interaction information, directly modeling the process of user interaction history to target items. 
 We have more clearly illustrated the differences between our motivation and existing diffusion-based methods in the Figure~\ref{fig:motivation}.

Consequently, obtaining the intermediate states required for diffusion models and inferring the sampling function that fits them presents a significant challenge. To address this, we introduce the Schrödinger Bridge~\cite{schrodinger1932theorie,leonard2013survey} into diffusion-based sequential recommendation, which considers how to find the transfer path with the minimum required cost given the initial and marginal distributions. On a technical level, the determination of a Schrödinger bridge capable of connecting two distributions is intricate. Therefore, we use a tractable Schrödinger bridge to simplify the process of establishing the connection and derive the sampling function from it, thus constructing our SdifRec model. Specifically, we first employ a Transformer model to process the historical interaction sequence, obtaining the current state representation of the user, which is considered as the initial distribution. The embedding of the targeted recommended item is regarded as the marginal distribution. Subsequently, we introduce a Schrödinger Bridge to establish the connection between these distributions, thereby eliminating the necessity of using Gaussian noise as the prior, a common practice in typical diffusion models. Furthermore, we design a connectivity model to reconstruct the representation of the target recommendation item at random moments. During the inference process, we initiate from the user's current state representation rather than Gaussian noise and iteratively apply the well-trained connectivity model to reveal the user's interests in the next moment. Finally, by computing and ranking the similarity between the user's next moment of interest and candidates, we recommend the target item.

Based on the propsoed SdifRec, the issues posed by prior constraints are effectively resolved.
Moreover, we have extended our focus to the respective strengths of SR and graph-based recommendation methods. SR can better model the dynamic evolution of user interests while the latter can more sensitively capture collaborative information between users and items. To combine the advantages of both forms, we propose an enhanced version of SdifRec called con-SdifRec. It utilizes user static representations obtained from pre-trained LightGCN to cluster users and uses the cluster information as conditional guidance for posterior distribution generation.
In summary, the main contributions of this paper include: 

\begin{itemize}[leftmargin=*]

\item
We are the first to introduce the Schrödinger Bridge into diffusion-based SR work, thereby presenting the SdifRec model. It directly models the connection between the user's current state and the target item, rather than relying on the conventional Gaussian distribution prior used in diffusion-based models.

\item
To capitalize on the strengths of both sequential recommendation and graph-based recommendation methods, we propose an extended version of SdifRec, termed con-SdifRec. It effectively utilizes collaborative information as conditional guidance to generate posterior distribution with extra information.

\item

We have conducted extensive experiments on three public benchmark datasets, comparing SdifRec with several state-of-the-art methods. The results have demonstrated significant improvements of SdifRec and con-SdifRec over baselines across various settings, verifying their efficiency and robustness.
\end{itemize}

\section{Related Work}
\subsection{Sequential Recommendation}
SR is a technique that suggests the subsequent item of potential interest, based on a user's historical interaction records~\cite{yin2023apgl4sr,shen2024exploring,yin2024entropy,han2024efficient, wu2024survey, han2023guesr}. This approach was initially implemented using techniques such as Markov Chain and Matrix Factorization~\cite{10.1145/2911451.2911489}. However, with the advent of neural networks, deep learning methods like GRU4Rec~\cite{hidasi2015session} have been employed to utilize Gated Recurrent Units (GRUs)~\cite{cho2014learning} to capture sequential dependencies within sequences of user behavior. Caser~\cite{tang2018personalized} and NextItNet~\cite{yuan2019simple} introduce Convolutional Neural Networks (CNNs)~\cite{lecun1998gradient} to learn local patterns in user behavior sequences. Graph neural networks (GNNs) have also gained attention for their ability to capture higher-order relationships among items like SR-GNN~\cite{DBLP:journals/corr/abs-1811-00855} and GCE-GNN~\cite{10.1145/3397271.3401142}. 
After Transformer~\cite{vaswani2017attention, wang2021hypersorec} appears, SASRec~\cite{kang2018self} is a pioneering work that introduces the architecture to the field of SR, becoming a mainstream framework. Additionally, BERT4Rec~\cite{sun2019bert4rec} draws inspiration from the BERT architecture and employs bidirectional encoders to capture bidirectional dependencies in sequences, using a masked language model to predict the user's next action.

In recent years, with the development of generative models, an increasing number of studies have begun to apply generative models such as VAE~\cite{vaswani2017attention} and GAN~\cite{creswell2018generative} to the field of SR, resulting in significant progress, such as MVAE~\cite{liang2018variational}, ACVAE~\cite{xie2021adversarial}, RecGAN~\cite{bharadhwaj2018recgan}, MFGAN~\cite{jung2020multi}. Nevertheless, models grounded in GANs typically necessitate adversarial training between the generator and discriminator. This process can often be unstable, leading to suboptimal performance~\cite{becker2022instability, nagarajan2017gradient}. Conversely, models founded on VAEs impose stringent assumptions about the posterior, which may constrain the quality of their generated hidden representations~\cite{kingma2016improved, sohl2015deep}. As a result, a few works in SR have turned their attention to the new paradigm of generative models - diffusion models~\cite{yang2023diffusion}. Among them, DiffuRec~\cite{li2023diffurec} and DiffRec~\cite{du2023sequential} directly apply diffusion models to the field of SR. DiffuASR~\cite{liu2023diffusion} utilizes user preference information as conditional guidance for personalized recommendations. DreamRec~\cite{yang2023generate} employs classifier-free guidance diffusion models to further leverage the conditional information of user preferences. Yet they are all troubled by the limitations imposed by the prior distribution.



\vspace{-2mm}
\subsection{Diffusion Models}

Diffusion models, inspired by non-equilibrium thermodynamics, have been introduced and demonstrate remarkable results in fields such as computer vision~\cite{li2022srdiff}, sequence modeling~\cite{li2022diffusion, tashiro2021csdi}, and audio processing~\cite{chen2020wavegrad, kong2020diffwave}. Currently, the mainstream diffusion models are mostly variations of the Denoising Diffusion Probabilistic Models (DDPM) by Ho et al~\cite{ho2020denoising}. and the Score-Based Generative Model (SGMs)~\cite{song2020score} proposed by Song et al. The latter uses Stochastic Differential Equations (SDE) to describe the data generation process, while DDPM can be seen as its discretized version with specific time step values. Given the broader conceptual framework of the SGMs, our subsequent discussions will be based on this form.

Classifier-guided diffusion is a subsequent work in diffusion models, mainly divided into classifier guided diffusion~\cite{dhariwal2021diffusion} and classifier-free guided diffusion~\cite{ho2022classifier}. The former requires training an additional classifier, and the quality of the classifier greatly affects the quality of the generated results. Classifier-free guided diffusion is an improvement on it, which constructs an implicit classifier gradient by discarding conditional information. This lays the foundation for subsequent work latent diffusion~\cite{rombach2022high}, which is a method of conducting diffusion processes in the latent space, thereby significantly reducing computational complexity. The use of classifiers also enables controllable generation~\cite{saharia2022photorealistic}, as demonstrated by the prominent work GLIDE~\cite{halgren2004glide}, which creates images based on textual descriptions.

\vspace{-2mm}
\subsection{Schrödinger Bridge}
The Schrödinger Bridge problem~\cite{schrodinger1932theorie} was first proposed in 1931, by Schrödinger, which is closely related to optimal control theory in mathematics~\cite{lefebvre2022entropy}, optimal transport problems\cite{chen2021optimal}, and the path integral methods in physics~\cite{kappen2005path}. Researchers like Valentin have applied the Schrödinger Bridge to Score-Based Generative Modeling using a method akin to Iterative Proportional Fitting~\cite{de2021diffusion}. They iteratively adjusted elements within the joint probability distribution to align with the target marginal distribution. Following this, Shi and others applied this method to path optimization problems~\cite{shi2023diffusion}. 


\vspace{-2mm}
\section{Preliminary}
\subsection{Score-Based Generative Model}
In this section, we will first introduce a Score-Based Generative Model (SGMs)~\cite{song2020score}, specifically a diffusion model represented in the form of Stochastic Differential Equations (SDEs). SGMs model the forward diffusion process using the stochastic differential equation:
\begin{equation}\label{eq1}
dx = f(x, t)dt + g(t)dw,  x_0 := {x}(0) \sim p_{0}=p_{\text {target}},
\end{equation}
where \( t \in [0, T] \), and \( w \) signifies Brownian motion, $p_{\text {target}}$ represents target distribution. The function \( f(\cdot, t): \mathbb{R}^d \rightarrow \mathbb{R}^d \)is a vector-valued function called the drift coefficient of $x(t)$, and \( g(\cdot): \mathbb{R} \rightarrow \mathbb{R} \) is a scalar function known as the diffusion coefficient of $x(t)$. The functions $f$ and $g$ determine the type of prior distribution $p_{\text{prior}}$ to which the forward process will diffuse, and they are typically designed to make the prior distribution a Gaussian distribution. As a remarkable result from Anderson (1982)~\cite{anderson1982reverse}, the reverse of the diffusion process is also a diffusion process, given by the following reverse-time SDE:
\begin{gather}
dx = [f(x, t) - g(t)^2 \nabla_x \log p_t(x)]dt + g(t)d\bar{w}, \label{eq2}\\
x_T := {x}(T) \sim p_{T} \approx p_{\text{prior}}, \notag
\end{gather}
where $\bar{w}$ is a standard Wiener process in reverse time. The term \( \nabla_x \log p_t(x) \), which represents the score function of the marginal density $p_t$, is the only unknown term in this reverse process. SGMs learns its approximate target \( s_{\theta}(x(t), t) \) through denoising score matching (DSM)~\cite{hyvarinen2005estimation}, with $s_{\theta}$ referred to as the denoising model:
\begin{align}\label{eq3}
\theta^* &= \arg\min_{\theta} \mathbb{E}_{t \sim U(0, T)}  \lambda(t) \mathbb{E}_{x(0)}\mathbb{E}_{x(t)|x(0)} \notag \\
&\quad \left[ \left\| s_{\theta}(x(t), t) - \nabla_x \log p_{0t}(x | x(0)) \right\|^2 \right].
\end{align}
Here, $\lambda(t)$ is a positive weighting coefficient, $t \sim \mathcal{U}(0, T)$. The joint distribution \( p_{0t}(x | x_0) \) is the conditional transition distribution from $x_0$ to $x(t)$, which is determined by the pre-defined forward SDE. To summarize, SGMs first utilize the diffusion process defined in Equation~(\ref{eq1}) to obtain the distribution $x(t)$ at intermediate time steps. Then, they minimize the loss defined in Equation~(\ref{eq3}) to train the denoising model $s_{\theta}$ and sample iteratively using the formula defined in Equation~(\ref{eq2}) to obtain the final result.

\vspace{-2mm}
\subsection{Schrödinger Bridge Problem}
We aim to obtain the corresponding intermediate state after replacing Gaussian noise with user interaction information as the prior distribution of the SGMs. To achieve this, we introduce the Schrödinger Bridge to model the process.
The Schrödinger Bridge (SB) problem is the optimization of path measures $p^{SB} \in \mathcal{P}(C)$ with constrained boundaries:
\begin{equation}\label{eq4}
    P^{SB} = \arg\min_{p} \{ KL(p||q) : p_0 = p_{\text {target}}, p_T = p_{\text {prior}} \},
\end{equation}
where $q \in \mathcal{P}(C)$ is a reference path measure. The above equation can be understood as finding a stochastic process with the minimum cost under the constraints of given initial and final state distributions. A common approach to solving Equation~(\ref{eq4}) is the Iterative Proportional Fitting (IPF)~\cite{fortet1940resolution} method:
\begin{equation}\label{eq5}
\begin{split}
\widetilde{p}_{2n+1} &= \arg\min_{\widetilde{p}} \{ KL(\widetilde{p}||\widetilde{p}_{2n}) : \widetilde{p}_T = p_{\text {prior}} \}, \\
\widetilde{p}_{2n+2} &= \arg\min_{\widetilde{p}} \{ KL(\widetilde{p}||\widetilde{p}_{2n+1}) : \widetilde{p}_0 = p_{\text {target}} \},
\end{split}
\end{equation}
with initialization $\widetilde{p}_{2n+1} = q$. Specifically, diffusion models can be viewed as the first even-iteration of Equation~(\ref{eq5}). 
Nevertheless, the method of obtaining the Schrödinger Bridge using IPF entails excessively high complexity. We will introduce the simplification operation in the next section.

\vspace{-2mm}
\subsection{Problem Definition}

Let $\mathcal{I}$ be the set of discrete items in the dataset, $\mathcal{U}$ be the set of users and $\mathcal{V}$ be the set of items.  For each user \({u} \in \mathcal{U}\)
, $ v_{1: n-1}=\left[v_{1}, v_{2}, \ldots, v_{n-1}\right]$ represents his historical interaction sequence sorted by timestamp. During the training process of a sequential recommender, learning involves maximizing the probability of the target item $v_{n}$, that is $p\left(v_{n} \mid v_{1}, v_{2}, \cdots, v_{n-1}\right) $. In the inference process, the generative sequential recommender predicts the probability of recommending the next item $v_{n+1}$ based on the entire sequence $\left[v_{1}, v_{2}, \cdots, v_{n}\right]$, that is $p\left(v_{n+1} \mid v_{1}, v_{2}, \cdots, v_{T-1}, v_{n}\right) $.

We utilize item embedding $e_j$ to signify the semantic representation of the latent features encapsulated within item $v_j$. So, \(e_{1:n-1}\) corresponds to the embedding of the historical interaction sequence, $e_n$ represents the embedding representation of the target item, which is also considered as the target distribution $x_0$ of our Schrödinger bridge. $h_u$ represents the hidden identifier of the current state of user $u$, which is also considered as the initial distribution $x_1$ of our Schrödinger bridge, Here let $T=1$. $x(t)$ Represents the state at time $t$ of the Schrödinger bridge connecting $x_0$ and $x_1$. In the following passage, we will use $x_1$ instead of Gaussian noise as the prior distribution of the diffusion model, and model the process from $x_1$ to $x_0$ using the Schrödinger Bridge.

\vspace{-2mm}
\section{Methodology of SdifRec}
In this section, we provide a detailed explanation of the proposed SdifRec, includes how to succinctly obtain the Schrödinger Bridge connecting the user's historical interactions and the recommended items, as well as how to obtain the corresponding sampling process. 

\begin{figure*}[htbp] 
\centering 
\includegraphics[width=0.95\textwidth]{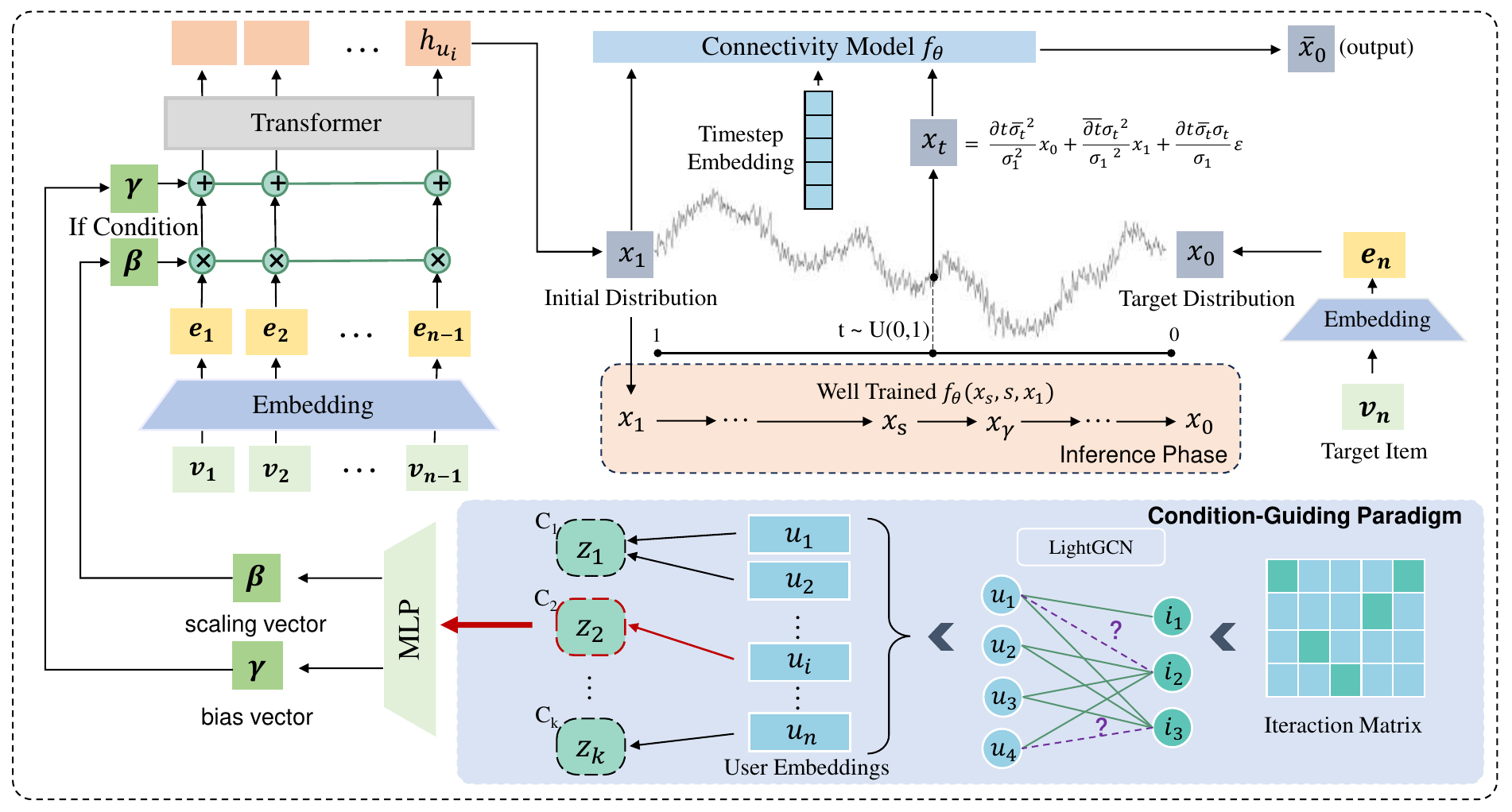} 
\vspace{-3.0mm}
\caption{Framework of SdifRec. The bottom left part of the figure illustrates the method of con-SdifRec.} 
\label{fig:structureSdifRec}
\label{fig:example} 
\vspace{-4.0mm}
\end{figure*}

\vspace{-2mm}
\subsection{Build the Schrödinger Bridge}
Initially, we input the embedded representation of the interaction sequence \(e_{1:n-1}=\left[e_{1}, e_{2}, \ldots, e_{n-1}\right]\) into a Transformer architecture similar to SASRec. To ensure its dimensionality matches that of the target item embedding $e_n$ (this is because we will subsequently compute their weighted sum), we select the last output from the Transformer as the current hidden state $h_u$ of the user. This process is represented by the approximator $h_\theta(\cdot)$:
\vspace{-2mm}
\begin{equation}\label{eq6}
    h_u = h_\theta(e_{1:n-1}).
    \vspace{-2mm}
\end{equation}

Next, we consider $h_u$ as the initial distribution $x_1$ of the Schrödin-ger bridge, and $e_n$ as the marginal distribution, which is our target distribution $x_0$. 
How to obtain the intermediate states of the Schrödinger bridge from the initial and target distributions remains a challenging task, and this stochastic process needs to satisfy Equation (\ref{eq1}). One feasible approach is to utilize the IPF method described in Equation (\ref{eq5}), which is a common method for solving the Schrödinger Bridge problem. However, it's worth noting that after parameter reparametrization, the traditional diffusion models require only one step to obtain the intermediate state from the initial state~\cite{ho2020denoising}. Therefore, employing the IPF method would significantly increase computational costs.

To deal with this, we consider the initial and target distributions as Gaussian distributions with specific means and variances and use a tractable Schrödinger Bridge model to simplify this problem. Specifically, we assume that the initial state follows the distribution $p_\text{initial} = \mathcal{N}(x_1, e^{2 \int_{0}^{1} f(\tau)d\tau}I)$, and the target state follows the distribution $p_\text{target} = \mathcal{N}(x_0, e^{2I})$. So, solving the Schrödinger Bridge by Equation~(\ref{eq5}) to satisfy Equation (\ref{eq1}) can be represented by the following partial differential equation:
\begin{equation}\label{eq7}
\left\{\begin{array}{l}
\frac{\partial \Psi}{\partial t}=-\nabla_{{x}} \Psi^{\top} {f}-\frac{1}{2} \operatorname{Tr}\left(g^{2} \nabla_{{x}}^{2} \Psi\right) \\
\frac{\partial \widehat{\Psi}}{\partial t}=-\nabla_{{x}} \cdot(\widehat{\Psi} {f})+\frac{1}{2} \operatorname{Tr}\left(g^{2} \nabla_{{x}}^{2} \widehat{\Psi}\right)
\end{array}\right. ,
\end{equation}
s.t.  $\Psi_{0} \widehat{\Psi}_{0}=p_{\text{target}}$, $\Psi_{T} \widehat{\Psi}_{T}=p_{\text {initial}}.$
 Here we present a lemma.
\begin{lemma}\label{lem:example}
    The result of Equation (\ref{eq7}) can be obtained by:
\begin{equation}\label{eq8}
\begin{aligned}
\hat{\Psi}_t &= \mathcal{N}(\alpha_t a, (\alpha_t \sigma_0^2 + \alpha_t \sigma_t^2)I), \\
\Psi_t &= \mathcal{N}(\bar{a}b, (\alpha_t \sigma_0^2 + \alpha_t^{-2} \sigma_t^2)I),
\end{aligned}
\end{equation}

where  $t \in[0,1]$, that is, let $T = 1$, and the values of $a$, $b$, etc. can be obtained from the following equation:
\begin{equation}\label{eq9}
\begin{split}
{a} &= {x}_{0}+\frac{\sigma^{2}}{\sigma_{1}^{2}}\left({x}_{0}-\frac{{x}_{1}}{\alpha_{1}}\right), \\
{b} &= {x}_{1}+\frac{\sigma^{2}}{\sigma_{1}^{2}}\left({x}_{1}-\alpha_{1} {x}_{0}\right), \\
\sigma^{2} &= \epsilon^{2}+\frac{\sqrt{\sigma_{1}^{4}+4 \epsilon^{4}}-\sigma_{1}^{2}}{2},
\end{split}
\end{equation}
and
\begin{equation}\label{eq10}
\begin{aligned}
    \alpha_{t} &= e^{\int_{0}^{t} f(\tau) \, \mathrm{d} \tau}, 
    \bar{\alpha}_{t} = e^{-\int_{t}^{1} f(\tau) \, \mathrm{d} \tau}, \\
    \sigma_{t}^{2} &= \int_{0}^{t} \frac{g^{2}(\tau)}{\alpha_{\tau}^{2}} \, \mathrm{d} \tau, 
    \bar{\sigma}_{t}^{2} = \int_{t}^{1} \frac{g^{2}(\tau)}{\alpha_{\tau}^{2}} \, \mathrm{d} \tau.
\end{aligned}
\end{equation}
When  $\epsilon \rightarrow 0$, $\widehat{\Psi}_{t}^{\epsilon}$, $\Psi_{t}^{\epsilon}$  converge to: 
$\widehat{\Psi}_{t}=\mathcal{N}\left(\alpha_{t} {x}_{0}, \alpha_{t}^{2} \sigma_{t}^{2} \boldsymbol{I}\right)$, $\quad \Psi_{t}=\mathcal{N}\left(\bar{\alpha}_{t} {x}_{1}, \alpha_{t}^{2} \bar{\sigma}_{t}^{2} \boldsymbol{I}\right).$
\end{lemma}

\begin{proof}
Due to space limitations, we provide a brief proof here. 
According to Itô's lemma~\cite{ito1951formula}, it can be derived that for the SDE satisfying Equation~(\ref{eq1}), there is
\begin{equation}
    {d} \left( \frac{x(t)}{\alpha_t} \right) = \frac{g(t)}{\alpha_t} {d}w,
\end{equation}
which leads to the result
\begin{equation}
    \frac{x(t)}{\alpha_t} - \frac{x_0}{\alpha_0} \sim \mathcal{N} \left( 0, \int_{0}^{t} \frac{g^2(\tau)}{\alpha^2_\tau} {d}\tau \mathcal{I} \right),
\end{equation}
then we conclude that \(\widehat{\Psi}_{t|0}(x(t)|x_0 = \mathcal{N} (\alpha_t x_0, \alpha^2_t \sigma^2_t \mathcal{I})\).

On the other hand, we can let \(s = 1 - t\) and conduct similar derivations for \(\Psi\), which finally leads to the result \(\Psi_{t|1}(x(t)|x_1 = \mathcal{N} (\bar{\alpha}_t x_1, \alpha^2_t \bar{\sigma}^2_t \mathcal{I})\). Then
\begin{equation}
    p_{\text{data}} = \widehat{\Psi}_0 \Psi_0 = \mathcal{N}(x_0, \epsilon^2 \mathcal{I}), \quad p_{\text{prior}} = \widehat{\Psi}_1 \Psi_1 = \mathcal{N}(x_1, \alpha_1^2 \epsilon^2 \mathcal{I}).
\end{equation}
We parameterize them as follows:
\begin{equation}
    \widehat{\Psi}_0 = \mathcal{N}({a}, \sigma^2 \mathcal{I}), \quad \Psi_1 = \mathcal{N}({b}, \alpha_1^2 \sigma^2 \mathcal{I}).
\end{equation}
Since the conditional transitions $\widehat{\Psi}_{t|0}, \Psi_{t|1}$ are known Gaussian, the marginals at any $t \in [0, 1]$ are also Gaussian :
\begin{equation}
    \widehat{\Psi}_t = \mathcal{N}(\alpha_t {a}, (\alpha_t^2 \sigma^2 + \alpha_t^2 \sigma_t^2) \mathcal{I}), \quad \Psi_t = \mathcal{N}(\bar{\alpha}_t {b}, (\alpha_t^2 \bar{\sigma}_t^2) \mathcal{I}).
\end{equation}

Then we can solve the coefficients ${a}, {b}, \sigma$ by boundary conditions.

\end{proof}

Then, according to the Lemma~\ref{lem:example}, the marginal distribution  $p_{t}=\widehat{\Psi}_{t} \Psi_{t}$  of the SB has a tractable form:
\begin{equation}\label{eq11}
    p_{t}=\Psi_{t} \widehat{\Psi}_{t}=\mathcal{N}\left(\frac{\alpha_{t} \bar{\sigma}_{t}^{2} {x}_{0}+\bar{\alpha}_{t} \sigma_{t}^{2} {x}_{1}}{\sigma_{1}^{2}}, \frac{\alpha_{t}^{2} \bar{\sigma}_{t}^{2} \sigma_{t}^{2}}{\sigma_{1}^{2}} \boldsymbol{I}\right).
\end{equation}
 That is, the intermediate state can be obtained by:
\begin{equation}\label{eq12}
    {x}_{t}=\frac{\alpha_{t} \bar{\sigma}_{t}^{2}}{\sigma_{1}^{2}} {x}_{0}+\frac{\bar{\alpha}_{t} \sigma_{t}^{2}}{\sigma_{1}^{2}} {x}_{1}+\frac{\alpha_{t} \bar{\sigma}_{t} \sigma_{t}}{\sigma_{1}} {\epsilon}, {\epsilon} \sim \mathcal{N}(\mathbf{0}, \boldsymbol{I}),
\end{equation}
This also reveals why we mentioned earlier that $h_u$ needs to have the same dimension as $e_n$, as inferring intermediate states involves multiplying them by separate coefficients and then adding them together.
For the definition of the drift coefficient $f$ and the diffusion coefficient $g$ that define the stochastic process, we provide two different definitions here:
\begin{align}
\begin{split}\label{eq13}
    \textbf{gmax: }  f(t) = 0, \quad g^{2}(t) = \beta_{0}+t(\beta_{1}-\beta_{0}).
\end{split} \\
\begin{split}\label{eq14}
    \textbf{VP: }  f(t) = -\frac{1}{2}(\beta_{0}+t(\beta_{1}-\beta_{0})), \quad g^{2}(t) = \beta_{0}+t(\beta_{1}-\beta_{0}).
\end{split}
\end{align}
The gmax form lacks a bias coefficient, indicating its suitability for modeling stochastic processes between two distributions with identical means, which is the opposite for VP. We can further analyze the intrinsic properties of SR through performance defined in two different forms in subsequent experiments.
The specific values corresponding to the two definitions for $\alpha_{t}, \bar{\alpha}_{t}, $ $\sigma_{t}, \bar{\sigma}_{t}$ can also be obtained from Equation (\ref{eq10}). 
So far we have successfully formulated and established the Schrödinger Bridge within the diffusion framework. Subsequently, we will elucidate this framework's training and inference processes within the SR domain.

\vspace{-2mm}
\subsection{Model Training}
In the training phase of the Schrödinger Bridge, we build our connectivity model $f_{\theta}$ with inputs initial distribution $x_1$, intermediate time distribution $x_t$ and time embedding $t$ to reconstruct the target distribution $x_0$. Since we have already used a transformer model to obtain the user's current state vector $h_u$, we define $f_{\theta}$ as a simple MLP. In terms of details, based on our experiments, we have observed that directly providing $x_t$ may lead to the model overly relying on the latent $x_0$ within $x_t$, resulting in suboptimal performance. Therefore, we introduce a parameter ${\alpha}$, which follows a normal distribution with mean and variance specified by hyperparameters ${\mu}$ and ${\sigma}$. We then multiply element-wise between $x_t$ and ${\alpha}$ before feeding it into the model. Therefore, the approximate value of $x_0$ obtained using the connectivity model is:
\begin{equation}\label{eq15}
    \hat{{x}}_{0} = f_{\theta}\left({\alpha} \odot {x}_{t}, t, {x}_{1}\right) .
\end{equation}

In practice, to allow the model to better learn the importance of time steps, we amplify $t$ by an amplification factor of $\lambda$, where $\lambda$ is a hyperparameter. Furthermore, due to better compatibility with cross-entropy loss and its suitability for SR~\cite{mao2023cross}. We have discarded the more commonly used loss function resembling Mean Squared Error (MSE) in the diffusion model. Instead, we have redefined the loss function as follows:
\begin{equation}\label{eq16}
    \mathcal{L}_{C E}=-\frac{1}{|\mathcal{U}|} \sum_{i \in \mathcal{U}} \log \left(\frac{\exp (\hat{{x}}_{0} \cdot {e}_{n})}{\sum_{i \in \mathcal{V}} \exp (\hat{{x}}_{0} \cdot {e}_{i})}\right) .
\end{equation}
The detailed training process can be found in Algorithm~\ref{alg:sdifrec}.
The well-trained connectivity model $f_{\theta}$, assists us in generating an approximate value for $x_0$ to be used in the subsequent sampling process during the model inference stage.

\subsection{Model Inferencing}
In the inference phase of the Schrödinger Bridge, our goal is to iteratively generate the target item embedding from the initial distribution $x_1$ obtained from the historical interaction sequence. The sampling process is performed using the following equation at time $s \in [0,1]$:
\begin{equation}\label{eq17}
\begin{split}
{x}_{t} &= \frac{\alpha_{t} \sigma_{t}^{2}}{\alpha_{s} \sigma_{s}^{2}} {x}_{s} + \alpha_{t}\left(1-\frac{\sigma_{t}^{2}}{\sigma_{s}^{2}}\right) {f}_{\theta}\left({x}_{s}, s, {x}_{1}\right) \\
&\quad + \alpha_{t} \sigma_{t} \sqrt{1-\frac{\sigma_{t}^{2}}{\sigma_{s}^{2}}} {\epsilon}, \quad {\epsilon} \sim \mathcal{N}(\mathbf{0}, \boldsymbol{I})
\end{split} ,
\end{equation}

\begin{equation}\label{eq18}
\begin{split}
{x}_{t} &= \frac{\alpha_{t} \sigma_{t} \bar{\sigma}_{t}}{\alpha_{s} \sigma_{s} \bar{\sigma}_{s}} {x}_{s} + \frac{\alpha_{t}}{\sigma_{1}^{2}}\left[\left(\bar{\sigma}_{t}^{2}-\frac{\bar{\sigma}_{s} \sigma_{t} \bar{\sigma}_{t}}{\sigma_{s}}\right) {f}_{\theta}\left({x}_{s}, s, {x}_{1}\right)\right. \\
&\quad \left.+\left(\sigma_{t}^{2}-\frac{\sigma_{s} \sigma_{t} \bar{\sigma}_{t}}{\bar{\sigma}_{s}}\right) \frac{{x}_{1}}{\alpha_{1}}\right].
\end{split}
\end{equation}

\vspace{-1.5mm}

Equations (\ref{eq17}) and (\ref{eq18}) correspond to the SDE and ODE (similarly proposed in the work of Song~\cite{song2020score}) forms of the Schrödinger Bridge model, where $t \in [0,s]$, $f_{\theta}$ is the connectivity model introduced in Section 4.3. By selecting the appropriate time steps and iteratively running the above process until $t = 0$, we can obtain the target item embedding $x_0$. In our work, we opted for a relatively simple uniform sampling of time. In fact, there are multiple ways to choose the time schedule, and we leave the exploration of different approaches for future work. Finally, the recommendation list is generated by selecting the $K$ items from the item set $\mathcal{V}$ that are closest to the target item embedding $x_0$.

So far, we have completed the introduction of the main part of SdifRec. Next, we will present the cluster center guidance paradigm that we have proposed for SdifRec, namely con-SdifRec, which is an effective improvement.

\vspace{-1.5mm}
\section{Method of Condition-guiding}
In the domain of SR,  user's historical interaction sequences serve as inputs to model various user dynamic behaviors.
In contrast, some graph-based recommendation methods excel at extracting collaborative information between users and items.
Consequently, in this section, our proposed method, con-SdifRec, inspired by classifier-free guided diffusion, tends to cluster collaborative user information and integrate it as a guiding condition for the sampling process, aiming  to constraint the user with Group homogeneity, thus harnessing both of these information types simultaneously.

\vspace{-2mm}
\subsection{Introducing Conditional Information}
To gain collaborative user information, we start by using a pre-trained LightGCN to obtain static user representations ${u_1, u_2, ..., u_{|\mathbf{U}|}}$. Subsequently, we further enhance the quality of collaborative information through clustering. Specifically, we initialize $k$ cluster centers $\{z_1, z_2, ..., z_k\}$, which will be jointly optimized during the training process. We calculate the cosine similarity $cosine(u_i, z_j) = u_i^{\top} z_j /\left(\|u_i\|_{2}\|z_j\|_{2}\right)$ between users and each cluster center to assign user clustering information. which is represented as a one-hot encoding $c_i$, i.e.
\vspace{-2mm}
\begin{equation}\label{eq19}
   c_i[j] = 
\begin{cases} 
1 & \text{if } j = \arg\max\limits_{l\in\{1,2,\dots,k\}} cosine(u_i, z_l), \\
0 & \text{otherwise}.
\end{cases}
\end{equation}

Subsequently, we provide $c_i$ as conditional guidance information to con-SdifRec, enabling the computation of the user's current state $h_{u_i}^{con}$ enriched with collaborative information. We need to modify the approximate function $h_\theta$ mentioned in Section 4.2 so that it can receive conditional information. To achieve this, we pass the conditional vector through an MLP layer to obtain scaling vector $\beta$ and bias vector $\gamma$, them input $\beta \odot e_{1:{n-1}} + \gamma$ into the transformer, and to differentiate, we designate the approximate function capable of receiving conditional information as $\hat{h}_\theta$:
\begin{equation}\label{eq20}
\begin{split}
h_{u_i}^{con} &= \hat{h}_\theta(e_{1:n-1}, c_i) \\
= & {\rm Transformer}( \beta \odot e_1 + \gamma, \beta \odot e_2 + \gamma, \ldots, \beta \odot e_{n-1} + \gamma).
\end{split}
\end{equation}
We also require an additional unconditional model:
\begin{equation}\label{eq21}
\begin{split}
h_{u_i}^{uncon} &= \hat{h}_\theta(e_{1:n-1}, \emptyset) = {\rm Transformer}( e_1, e_2, \ldots, e_{n-1}).
\end{split}
\end{equation}
The remaining forward process remains the same as SdifRec, where we still use $h_{u_i}^{con}$ or $h_{u_i}^{uncon}$ as $x_1$, which serves as the input for the connectivity model $f_{\theta}$ to obtain the target distribution with conditional information $\hat{x}_0$. 

\subsection{Joint Training and Conditional Sampling}
During the training phase, we have referenced the form of classifier-free guided diffusion and jointly trained models with and without conditional information. Specifically, we use $h_{u_i}^{uncon}$ without conditional information as the input for the connectivity model $f_{\theta}$ with a probability of $p$. Conversely, with a probability of $1$-$p$, we utilized $h_{u_i}^{con}$, which incorporates conditional information, as input for the connectivity model.

During the sampling phase, we use the hyperparameter $w$ to control the strength of the influence of guidance signal $c_i$, and replace ${f}_{\theta}\left({x}_{s}, s, {x}_{1} = h_{u_i}\right)$ with 
\begin{equation}\label{eq22}
    \tilde{f}_{\theta}\left({x}_{s}, s, h_{u_i}, h_{u_i}^{con}\right)=(1+w) f_{\theta}\left({x}_{s}, s, h_{u_i}^{con}\right)-w f_{\theta}\left({x}_{s}, s, h_{u_i}^{uncon}\right)
\end{equation}
to complete Equation (\ref{eq13}) or (\ref{eq14})'s sampling process.
The rest remains the same as SdifRec, and we still obtain the final target embeddings through iterative sampling. Overall, we illustrated our SdifRec and con-SdifRec in Figure~\ref{fig:structureSdifRec}.

It is worth noting that here we provide the form of conditional guidance, and the available conditions are not limited to using clustering conditions obtained from fixed user representations. Multimodal information such as text embeddings and other side information can also serve as guidance conditions, which we leave for future research on diffusion-based SR. 

\begin{table}[t]
\centering
\caption{The detailed description and statistics of datasets}
\vspace{-2mm}
\label{tab:dataset-comparison}
\begin{tabular}{L{0.8cm}C{0.7cm}C{0.8cm}C{1.6cm}C{1.3cm}C{1.2cm}}
\toprule
\textbf{Dataset} & \textbf{\#Seq} & \textbf{\#items} & \textbf{\#Interactions} & \textbf{Avg.length} & \textbf{Sparsity}\\ \midrule
Beauty & 22,363 & 12,101 & 198,502 & 8.53 & 99.93\% \\
Toys & 19,412 & 11,924 & 167,597 & 8.63 & 99.93\% \\ 
Yelp & 30,983 & 29,227 & 321,087 & 10.30 & 99.96\%\\
\bottomrule
\end{tabular}
\vspace{-5mm}
\end{table}
\section{Experiment}
\begin{table*}[ht]
\centering
\caption{Experimental results(\%) of our SdifRec and other baseline models on three datasets. The best results are highlighted in bold, the second-best results are underlined, and * indicates significant improvements relative to the best baseline (t-test P<.05), with the relative improvements denoted as Improv.}
\vspace{-2mm}
\label{tab:experimental-results}
\begin{tabular}{@{}lccccccccccc@{}}
\toprule
Dataset & Metric & GRU4Rec & Caser & SASRec & BERT4Rec & ACVAE & MFGAN & DiffuRec & DreamRec & \textbf{SdifRec} & Improv. \\ \midrule
\multirow{4}{*}{Beauty} & HR@5  & 1.9421 & 2.6045 & 3.3372 & 2.4384 & 3.3167 & 3.1521 & \underline{5.5758} & 4.9816 & \textbf{6.0915}*& 9.25\% \\
                        & HR@10 & 2.9257 & 4.1920 & 6.3492 & 3.1205 & 6.2487 & 6.0017 & \underline{7.9068} & 6.9814 & \textbf{8.1943}* & 3.64\% \\
                        & NDCG@5 & 1.4234 & 1.2321 & 2.3741 & 1.6534 & 2.3941 & 2.2154 & \underline{4.0047} & 3.2145 & \textbf{4.3671}* & 9.05\% \\
                        & NDCG@10 & 1.8952 & 2.5021 & 3.2174 & 2.0167 & 3.2025 & 3.1645 & \underline{4.7494} & 3.9712 & \textbf{5.0664}* & 6.67\% \\
\bottomrule
\multirow{4}{*}{Toys}& HR@5  & 1.9565 & 1.8684 & 4.3219 & 2.2984 & 3.0987 & 2.5976 & \underline{5.5650} & 5.1044 & \textbf{5.8826}* & 5.71\% \\
                        & HR@10 & 2.8682 & 2.7985 & 6.5984 & 2.9948 & 5.5632 & 5.1952 & \underline{7.4587} & 6.3497 & \textbf{7.5844} & 1.69\% \\
                        & NDCG@5 & 1.3684 & 1.0651 & 2.9268 & 1.1659 & 2.0986 & 1.8287 & \underline{4.1667} & 3.1621 & \textbf{4.4730}* & 7.35\% \\
                        & NDCG@10 & 1.8461 & 1.6984 & 3.4682 & 1.5068 & 2.9463 & 2.2068 & \underline{4.7724} & 3.9117 & \textbf{4.9773}* & 4.29\% \\
\bottomrule
\multirow{4}{*}{Yelp} & HR@5  & 1.6142 & 1.6865 & 1.6213 & 1.8964 & \underline{1.9546} & 1.8974 & 1.4195 & 1.7351 & \textbf{2.3302}* & 19.22\% \\
                        & HR@10 & 2.9740 & 2.9986 & 3.1074 & 3.2468 & \underline{3.4685} & 3.3552 & 1.5497 & 1.9254 & \textbf{3.7519}* & 8.17\% \\
                        & NDCG@5 & 0.9986 & 0.9465 & 0.9627 & 1.1086 & \underline{1.2527} & 1.1865 & 1.2844 & 1.1742 & \textbf{1.5744}* & 25.7\% \\
                        & NDCG@10 & 1.2985 & 1.3786 & 1.3624 & 1.3889 & \underline{1.5854} & 1.4652 & 1.3268 & 1.5177 & \textbf{2.0178}* & 27.3\% \\

\bottomrule
\end{tabular}

\vspace{-3mm}
\end{table*}

\subsection{Experiment Settings}
\subsubsection{Datasets}
We selected three real-world datasets widely used in the sequential recommendation to evaluate the performance of our SdifRec: \textbf{Amazon Beauty} and \textbf{Amazon Toys} are two subcategories of the Amazon \footnote{https://cseweb.ucsd.edu/~jmcauley/datasets/amazon\_v2/} dataset~\cite{mcauley2015image}, encompassing data collected from May 1996 to July 2014 on the Amazon online store. 
\textbf{Yelp} \footnote{https://www.yelp.com/dataset}~\cite{yelp} is a large-scale social media and business review dataset widely used for research and development. Detailed descriptions and statistics for these datasets are provided in Table~\ref{tab:dataset-comparison}.

\vspace{-1mm}
\subsubsection{Baselines}
We compared SdifRec with eight state-of-the-art sequential recommendation methods, including four \textit{conventional sequential methods} and four \textit{generative sequential methods}:

\noindent The four \textit{conventional sequential methods} include:
\begin{itemize}[leftmargin=*,align=left]
\item \textbf{GRU4REC}\cite{hidasi2015session} is a classical RNN-based sequential recommendation model with a Gated Recurrent Units.
\item \textbf{Caser}\cite{tang2018personalized} applies CNN with vertical and horizontal convolutional layers to capture long and short-term user preferences.
\item \textbf{SASRec}\cite{kang2018self} utilizes a causal Transformer architecture with a self-attention mechanism to model sequential user behavior.
\item \textbf{BERT4REC}\cite{sun2019bert4rec} proposes a bidirectional Transformer with a cloze task predicting the masked target items for SR.
\end{itemize}

\noindent The four \textit{generative sequential methods} include:

\begin{itemize}[leftmargin=*,align=left]
\item \textbf{ACVAE}\cite{xie2021adversarial} proposes an adversarial and contrastive variational autoencoder for SR combining the ideas of CVAE and GAN.
\item \textbf{MFGAN}\cite{jung2020multi} utilizes multi-factor generative adversarial network(GAN) to consider information from various factors.
\item \textbf{DiffuRec}\cite{li2023diffurec} introduces the diffusion model into the field of SR reconstructing target item representation from a Transformer backbone with the user’s historical interaction behaviors.
\item \textbf{DreamRec}\cite{yang2023generate} uses the historical interaction sequence as conditional guiding information for the diffusion model to enable personalized recommendations.
\vspace{-1mm}
\end{itemize}

\subsubsection{Evaluation Protocols}
Following the previous work~\cite{kang2018self,jung2020multi,yang2023generate}, we employ the leave-one-out strategy for performance evaluation across all datasets. Concretely, we consider the last interaction as the test set, the second-to-last interaction as the validation set, and all preceding interactions as the training set. We evaluate all models using metrics HR@K (Hit Rate) and NDCG@K (Normalized Discounted Cumulative Gain) and report experimental results for $K = {5, 10}$. Here, HR@K measures the proportion of hits among the top $K$ recommended items, and NDCG@K provides further evaluation of ranking performance by considering the positions of these hits in the ranking list. We rank all candidate items for target item prediction~\cite{krichene2022sampled}
.

\subsubsection{Specific Implementation Details}
We present the details of SdifRec and con-SdifRec below. For SdifRec, we set the dropout rate to 0.2 for the embedding layer. The embedding dimension and hidden layer dimension were both set to 128. We made two attempts for functions $f$ and $g$ as gmax and VP shown in Equations (\ref{eq13}) and (\ref{eq14}), where $\beta_0$ was set to 0.01, and $\beta_1$ varied within the range of [10, 20, 30, 40, 50]. Values for $\mu$ and $\sigma$ were selected from [0.001, 0.01, 0.1, 1], and the learning rate was set to 0.001. We initialized the parameters of the Transformer using Xavier normalization distribution and set the number of blocks to 4. We explored the sampling steps in the range of [10, 15, 20, 25, 28, 30, 32, 35] and compared both SDE and ODE sampling methods. For con-SdifRec, we obtain user embeddings pre-trained using a loss function based on DirecteAU for LightGCN, and attempted clustering centers $k$ in the range of [5, 7, 10, 12, 15, 17, 20]. For the guidance strength $w$, we tried [0.3, 0.5, 0.8, 1.0, 1.3, 1.5, 1.8, 2.0, 3.0]. To ensure the fairness of our experiments, we optimal all baselines according to the original paper, and repeat 10 times for a more stable evaluation.

\vspace{-2mm}
\subsection{Overall Performance}
In this section, we compared SdifRec with baseline models in terms of top-K recommendation performance, and the results are summarized in Table~\ref{tab:experimental-results}
. We can draw the following observations: 

1). SdifRec has achieved significant improvements on all three datasets, demonstrating that the Schrödinger Bridge Diffusion model can obtain effective item representations and reasonably model the connection between the user's current state(derived from the history of interaction sequences) and the target recommended items. This is also demonstrated by the large gap between SASRec and our model since removing the Schrödinger Bridge from SdifRec can be seen as somewhat similar to SASRec.

2). Methods based on generative models, namely ACVAE, MFGAN, DiffuRec, and DreamRec, generally perform well on different datasets, outperforming traditional sequential recommendation algorithms. This validates that generative models can help us obtain good hidden representations of items and users. Among the methods based on generative models, Diffusion-based models DiffuRec and DreamRec tend to perform better overall than ACVAE and MFGAN. 
We believe this is because the diffusion model does not suffer from the issue of posterior collapse, which VAE and GAN may face, where the generated hidden representations may contain little information about users and items.

\vspace{-2mm}
\begin{figure}[htbp]
  \centering
  \includegraphics[width=1\columnwidth]{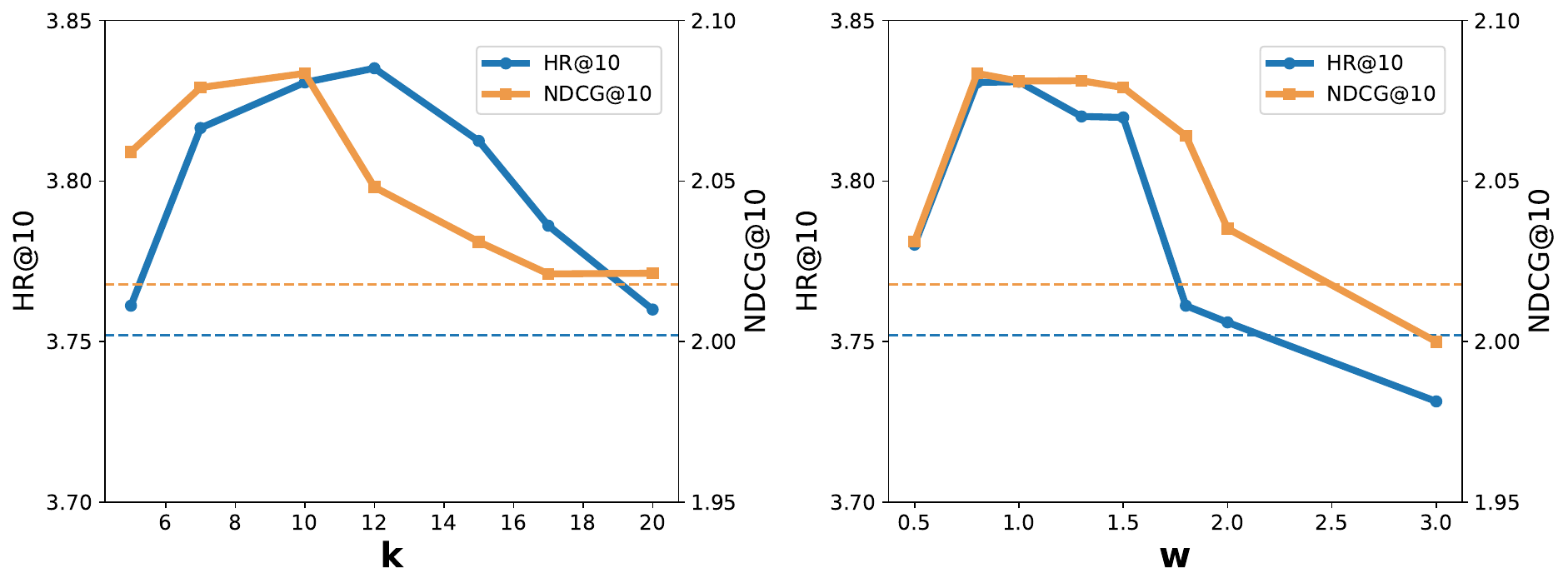}
\vspace{-7mm}
  \caption{The influence of different numbers of clustering centers and guidance strength on the results of con-SdifRec.}
  \label{fig:cluser_centers}
  
\vspace{-4mm}
\end{figure}

3). 
While DiffuRec remains the best-performing baseline on the Beauty and Toys datasets, only surpassed by SdifRec, it fails to deliver satisfactory results on the Yelp dataset. In contrast, SdifRec continues to demonstrate exceptional performance. This suggests that our model has successfully mitigated the problem of information loss, thereby offering a substantial advantage when handling large and intricate datasets. This also indicates that using the user's historical interaction information as a prior distribution is more effective than treating it as conditional information, highlighting the importance of replacing the Gaussian distribution in diffusion. 
Additionally, we chose to use SGMs simply to provide a more unified framework for theoretical development and modification. This decision was made because DDPM can be considered a specific case of SGMs under certain sampling conditions.
Furthermore, our experiments revealed that the sampling method for time steps did not significantly impact the model results. Therefore, in the context of SR, there is not a substantial difference between DDPM and SGMs. Although SdifRec and DiffuRec are based on SGMs and DDPM, respectively, the improvement in performance primarily stems from the introduction of the Schrödinger Bridge.

\vspace{-2mm}
\subsection{Analysis of con-SdifRec}
To ascertain the positive impact of incorporating collaborative information into SdifRec on recommendation performance, we compared the performance of con-SdifRec with SdifRec on three datasets. The results are shown in Table~\ref{tab:epsilon-elbo}. Experimental results indicate that con-SdifRec exhibits improvements across all three datasets compared to SdifRec. This suggests that con-SdifRec can leverage the advantages of both matrix-based methods and sequential recommendation methods simultaneously.

\begin{table*}[htbp]

\vspace{-2mm}
\centering
\caption{Comparison of the performance between con-SdiRec and SdifRec. “R@K” is short for “HR@K” and “N@K” is short for “NDCG@K”. Bold indicates better performance.}

\vspace{-3mm}
\begin{tabular}{@{}lcccccccccccc}
\toprule
& \multicolumn{4}{c}{Beauty} & \multicolumn{4}{c}{Toys} & \multicolumn{4}{c}{Yelp} \\ 
\cmidrule(r){2-5} \cmidrule(r){6-9} \cmidrule(r){10-13}
& R@5 & R@10 & N@5 & N@10 & R@5 & R@10 & N@5 & N@10 & R@5 & R@10 & N@5 & N@10 \\
\midrule
SdifRec & 6.0915 & 8.1943 & 4.3671 & 5.0664 & 5.8826 & 7.5844 & 4.4730 & 4.9773 & 2.3302 & 3.7519 & 1.5744 & 2.0178 \\
con-SdifRec& \textbf{6.2560} & \textbf{8.3664} & \textbf{4.4501} & \textbf{5.1879} & \textbf{6.0710} & \textbf{7.7891} & \textbf{4.6922} & \textbf{5.1465} & \textbf{2.4392} & \textbf{3.8307} & \textbf{1.6451} & \textbf{2.0835}\\

\midrule

improvement&
\textcolor{red}{+2.7\%} & \textcolor{red}{+2.1\%} & \textcolor{red}{+1.9\%} & \textcolor{red}{+2.4\%} & \textcolor{red}{+3.2\%} & \textcolor{red}{+2.7\%} & \textcolor{red}{+4.9\%} & \textcolor{red}{+3.4\%}  & \textcolor{red}{+4.7\%} & \textcolor{red}{+2.1\%} & \textcolor{red}{+4.5\%} & \textcolor{red}{+6.6\%}\\
\bottomrule
\end{tabular}
\label{tab:epsilon-elbo}

\vspace{-3mm}
\end{table*}

Furthermore, since the number of clustering centers $k$ and the clustering guidance strength $w$ have significant effects on con-SdifRec, we also carried out experiments on these two parameters. The experimental results with the dashed line to display the results of SdifRec for observation are shown in Figure~\ref{fig:cluser_centers}.
The figure illustrates that (1) optimal performance is attained when the number of clustering centers $k$ approximates 10. Furthermore, it is observed that altering the number of clusters within this normal range does not adversely affect the outcomes.
(2) The optimal performance is achieved when the guidance strength $w$ is set to around 0.8, and tremendous values of $w$ may even result in inferior performance compared to SdifRec. We believe this is because the guidance strength determines the degree to which the recommendation results converge towards the clustering centers, and excessively strong guidance may excessively rely on the clustering results.

\vspace{-4mm}
\begin{figure}[htbp]

  \centering
  \includegraphics[width=1\columnwidth]{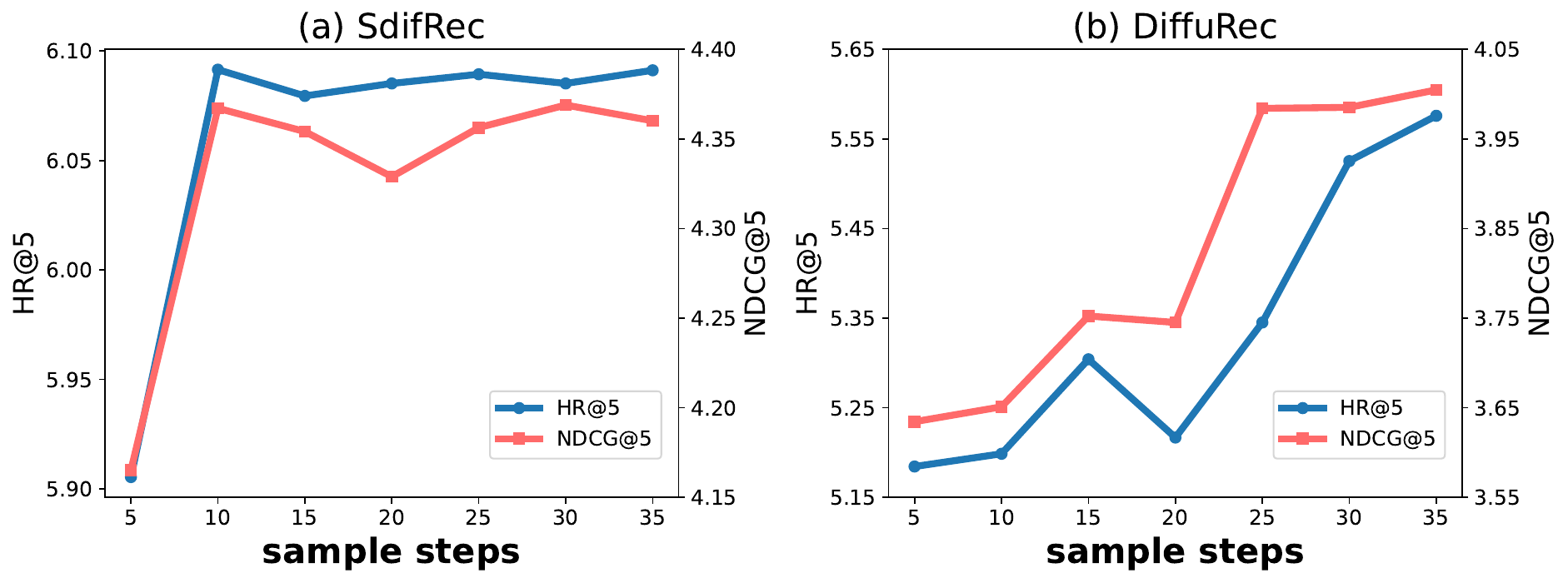}

  \caption{The impact of sample steps on SdifRec and DiffuRec.}
  \label{fig:sample_time}
\end{figure}
\vspace{-6mm}
\subsection{Efficiency Analysis}
We analyze the efficiency of SdifRec by examining two aspects: training time and inferencing time. We conducted a time comparison with DiffuRec, the most representative method in the baseline.
\vspace{-5mm}
\subsubsection{Analysis on Training Time}
Due to the adoption of similar structures between SdifRec and DiffuRec, the training time per epoch for our model is close to DiffuRec, at around 14 seconds each. However, our model exhibits superior performance in terms of convergence speed. 
For a fair comparison, we keep the same parameters between DiffuRec and SdifRec.
Specifically, after multiple repeated experiments, we found that our SdifRec converges in approximately 60-70 epochs on the Amazon Beauty dataset, while DiffuRec requires around 100 epochs. On the Toys dataset, we need 110-120 epochs to converge, whereas DiffuRec requires 180-190 epochs. This is because we no longer engage in the process of exploring data to noise, and the model doesn't need to use the user's current state as auxiliary information. Instead, it directly learns the embedding representation from the user's current state to the items to be recommended, which is a more direct process. Overall, the training time for SdifRec is reduced by more 30\% compared to DiffuRec, which is a significant improvement.

\vspace{-1.5mm}
\subsubsection{Analysis on Inferencing Time}
The number of sampling steps is the most important factor affecting the inference speed of diffusion models with similar architectures. As a conclusion, we only need 10-12 sampling steps to achieve optimal results, while DiffuRec requires 35-40 steps. This is because our improvement enables the sampling process to start from the user's current state rather than pure noise, resulting in faster inferencing. 
We have provided a detailed demonstration in Figure~\ref{fig:sample_time}.

\vspace{-2mm}
\subsection{Robustness Analysis}
\begin{figure}[htbp]
  \vspace{-2mm}
  \centering
  \includegraphics[width=0.95\columnwidth]{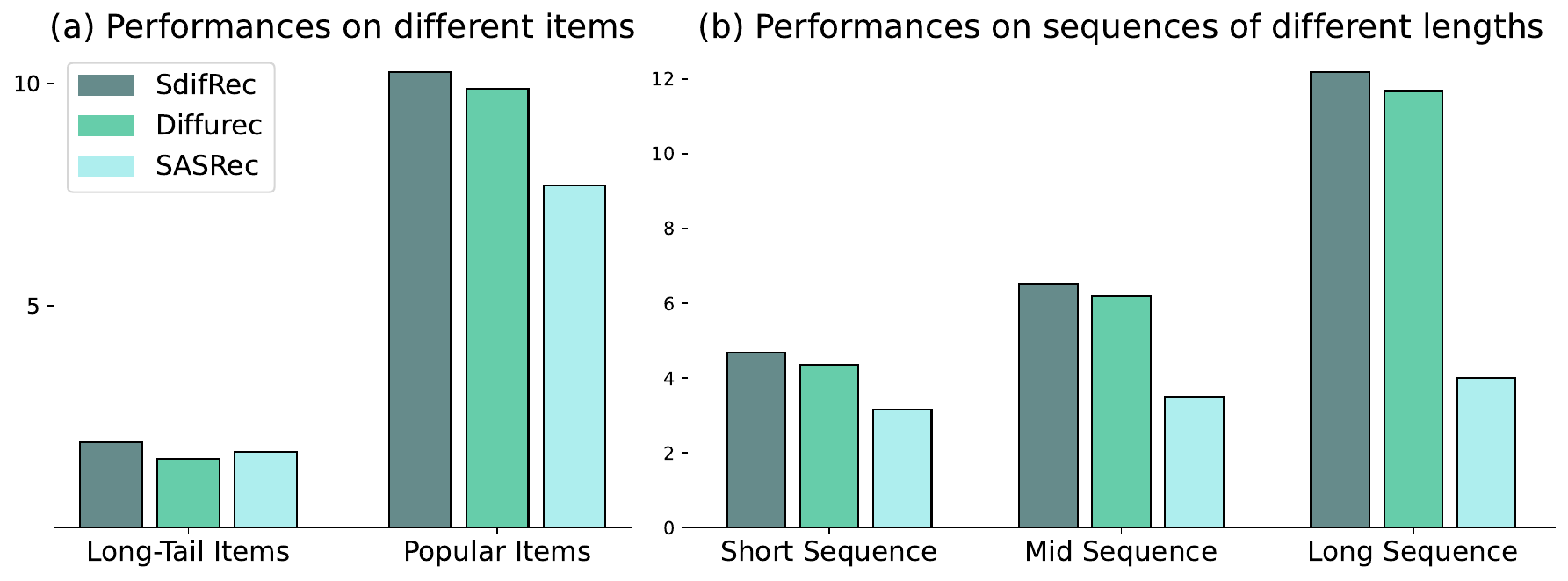} 
  \vspace{-3mm}
  \caption{Performance comparison between SdifRec, DiffuRec and SASRec under different items and sequence lengths on the Amazon Beauty dataset.}
  \label{fig:robustness}
\vspace{-5mm}
\end{figure}

We analyzed the impact of different sequence lengths and item popularity on the recommendation results of SdifRec and compared it with DiffuRec and SASRec to verify the robustness of SdifRec. We have presented the results on the beauty of Amazon in Figure~\ref{fig:robustness}.
Specifically, we consider the top 20\% most frequently occurring items as popular items, while the rest are categorized as long-tail items. It can be observed that all three models perform better when interacting with more frequently interacted items, indicating that increased interaction frequency helps the models learn more about these items. SdifRec also achieves the best performance for both long-tail and popular items, demonstrating the effectiveness of our model in general scenarios.

Regarding sequence length, we divided the sequences into short (0 to 5 inclusive), medium (6 to 10 inclusive), and long (greater than 10) based on their lengths. SASRec exhibits relatively small performance variations across different lengths, whereas Diffurec and our SdifRec show more significant improvements in handling long sequences. This suggests that the introduction of diffusion models and the Schrödinger bridge of SdifRec is beneficial for better capturing different length of sequences compared to directly using the Transformer architecture. 
Overall, our model performs the best across all settings, demonstrating the robustness of SdifRec.

\subsection{Impact of Different Settings}

We compared the differences in results brought about by different configurations, including (1) the choices of $f$ and $g$, (2) the sampling methods for SDE and ODE, (3) the setting of $\beta_1$.
We have listed the results on the Amazon dataset in Table~\ref{tab:diff_setting} from which it can be observed that the SDE method slightly outperforms the ode method, indicating its better suitability for SR tasks. Additionally, the random process defined by the gmax method performs significantly better than the VP method, while the performance of the VP method is poor. This suggests that recommendation tasks may be better suited for random processes without bias, likely due to the small differences in distribution means between the current user state and the items to be recommended. The significant impact of different formula settings on the Schrödinger bridge is evident, indicating the importance of further exploration in formula research. Regarding the setting of $\beta_1$, under the random process defined by gmax, it determines the diffusion level, and we found that setting it to 50 achieves the best performance, indicating that some level of noise disturbance contributes to the model learning better representations. Under the random process defined by VP, the bias coefficient is affected by 
$\beta_1$, and we found that setting it to only 20 achieves optimal results, further suggesting that modeling sequence recommendations does not require a significant amount of bias.

\vspace{-2mm}

\begin{table}[h]
\centering
\caption{The HR@10 metric under different configurations on the Amazon dataset.}
\label{tab:diff_setting}
\resizebox{\columnwidth}{!}{%
\begin{tabular}{@{}lcccccc@{}}
\toprule
\multirow{2}{*}{\textbf{Method}} &  & \multicolumn{5}{c}{Values of \( \beta_1 \)} \\ 
\cmidrule(lr){3-7} 
 &  & 10 & 20 & 30 & 40 & 50 \\ 
\midrule
\multirow{2}{*}{\textbf{gmax}} & \textbf{SDE} & 7.5561 & 7.7875 & 7.9061 & 8.0965 & 8.1943\\
 & \textbf{ODE} & 7.2524 & 7.4638 & 7.6052 & 7.8451 & 8.0145\\
\midrule
\multirow{2}{*}{\textbf{VP}} & \textbf{SDE} & 6.0142 & 6.1865 & 6.1653 & 5.8415 & 5.4653 \\
 & \textbf{ODE} & 5.5987 & 5.7653 & 5.6451 & 5.1652 & 4.6584 \\
\bottomrule
\end{tabular}%
}
\vspace{-1mm}
\end{table}

\vspace{-1mm}
\vspace{-3mm}
\section{Conclusion}

In conclusion, this paper introduced SdifRec, a novel framework introducing the Schrödinger bridge towards diffusion-based SR models. The Schrödinger bridge introduced into the diffusion model addressed the limitation of the prior distribution. Subsequently, we introduced the extended version of SdifRec, con-SdifRec, which effectively utilizes cluster information as conditional guidance, making effective usage of collaborative information. Extensive experiments and analysis on three benchmark datasets validated the effectiveness, efficiency, robustness, and stability of SdifRec and con-SdifRec.
In the future, we believe there are many promising ideas worth further exploration in this direction: 1) We find that the settings of different random processes and sampling methods have a significant impact on the recommendation results, so exploring forms more suitable for the recommendation domain is intriguing. 2) The paradigm of con-SdifRec with conditional guidance allows us to incorporate more modal information, and further exploration beyond clustering guidance may lead to additional improvements.


\bibliographystyle{ACM-Reference-Format}
\bibliography{sample-base}
\end{document}